\documentclass[preprint,12pt]{elsarticle}

\usepackage{amssymb}

\journal{Acta Astronautica}

\begin{document}
\newcounter{graf}
\begin{frontmatter}



\title{Using Interplanetary Medium as Propellant for Plasma Thruster Propulsion}
\author[a,b]{A.R. Karimov}
\author[b]{O.V. Yakovlev}
\author[c]{P.A. Murad}
\address[a]{Department of Electrophysical Facilities, National Research Nuclear University MEPhI, Kashirskoye shosse 31, Moscow, 115409, Russia}
\address[b]{Institute for High Temperatures, Russian Academy of Sciences, Izhorskaya 13/19, Moscow, 127412, Russia}
\address[c]{Morningstar Applied Physics, LLC, Vienna, USA}
\begin{abstract}
Consuming long-range interplanetary medium with fuel for driving plasma thrust is vital. Methods of capturing the space environment depending on its density and the ship velocity are indispensable issues to extract the space environment as a propellant. This matter of neutral fluxes would use solar radiation to create ionization in a plasma thruster. Extracting propellant at certain ranges look extremely promising where a low thrust device operates continuously to deliver a celestial target close at the speed of light. However, at longer distances, the density of space rapidly diminishes and it may be impossible to maintain at a plasma thruster's continuous operation mode but instead might be workable with a pulsed engine. Nonetheless, at still longer-range interstellar missions, the insufficient propellant will require examining other alternatives than a plasma thruster for other candidate propulsion system, possibly either propellantless or a propulsor is driven by electric and/or magnetic fields to satisfy these mission objectives.
\end{abstract}

\end{frontmatter}


\section{Introduction}

The exploration of the Solar system requires a creation of a space engine with a large specific impulse. Unfortunately, such characteristics can't be provided by traditional chemical reactive engines. This follows directly from the Tsiolkovski's formula:
\begin{equation} 
v_f=v_{ex}\ \ln\left(\frac{M_s+M_p}{M_s}\right)\/,
\label{fl_1}
\end{equation}
here $v_{ex}$ is the effective exhaust velocity, $M_{s}$ is the rocket mass, $M_p$ is the propellant mass, $v_f$ is the final velocity of the rocket. As an alternative to chemical engines, one can use nuclear or plasma-driven engines. End-Hall thrusters are capable of creating the exhaust velocity of the order $5\cdot 10^5$ m/s, the nuclear engine - $10^4$ m/s while the maximum velocity of chemical engines is only ${10}^3$ m/s \cite{SB}-\cite{GLK}. Moreover, the propellant weight requirements may become prohibitive in terms of cargo and
flight duration.

Nowadays there are several schemes of plasma thrusters: coaxial plasma thruster, stationary plasma thruster, Hall plasma thruster, plasma thruster energy/momentum exchange in crossed magnetic fields \cite{GK}-\cite{KM18}. These devices differ from each other in the way of creating acceleration to provide thrust and performance. Nevertheless, there is a generic feature for all devices: in order to work, one should obtain plasma from neutral atoms treated as propellants, and then one should accelerate this plasma flow to generate thrust. It is clear that this takes some energy and material which is continuously expended. In comparison with chemical thrusters, which can use only the reserve propellant, in plasma thrusters, one can also use solar energy and interplanetary medium as a fuel.

In this regard, it is also worth noting that the use of the interplanetary medium as a fuel for spacecraft has been discussed since Bussard \cite{Bus,BD} up to nowadays works (see, for example, Refs. \cite{Cas}-\cite{BG} and references therein) where this idea has been considered for nuclear thrusters. The interstellar engine (so-called Bussard ramjet) scoops hydrogen from the interstellar medium and uses this as both a fuel and energy source by way of a fusion reactor. However, as already mentioned, the exhaust velocity for a nuclear jet engine is less than the one for plasma thrusters. So it would be of interest to discuss such an approach to using plasma thrusters.

\section{Local interplanetary medium characteristics}

Before discussing these potential applications, it is worth considering basic properties of the local interplanetary medium in the Solar system.

As the interplanetary medium is a composition of substance and fields which fill space inside the Solar system, it appears justified to first study its components. The solar system is assumed to be in a low-density, high-temperature portion plasma of the interstellar medium. According to the information previously presented \cite{Pik,Val}, the interstellar magnetic field strength $B_{ism}$ is the order of a few ${10}^{-14}$ weber/cm$^{2}$. Ignoring the zero-point field and matter, the main components of the interplanetary system are the solar wind, the high energy charged particles, the interplanetary dust, and neutral gas
\cite{PS}. The solar wind includes the ionized flux of the hydrogen ions of density $n_H=5$ cm$^{-3}$ with velocity 250-750 m/s moving from the Solar Corona \cite{PS}. Evidently, there are particles of the flow, where it is impossible to register. Taking into consideration such unknown part of the flow, in the present estimations, we will slightly increase the density of the solar wind $n_w$ by setting it equal to 10 particles per cm$^{-3}$.

A source of small particles forming a dust component is from the collapsing nucleus of comets and the collisions of different bodies in the asteroid belt. Of these, the smallest particles have gradually moved to the Sun under the Poynting-Robertson effect \cite{PS,Bei}. The physical meaning of this effect is as follows. Sunlight might affect the moving particles but the direction of this action does not coincide with the line directed from the Sun to the particle. Such a deviation is caused by  light aberration. In this situation, the sunlight action is directed against the velocity of the particle, and therefore it slows down the particle motion. As a result, in the space of the solar system between the Earth and the Sun, there exists an extremely uneven distribution of dust. Its main quantity is concentrated in the ecliptic plane (sometimes referred to as the zodiacal dust cloud). The dust concentration in the zodiac cloud decreases with distance from the Sun and from the plane of the ecliptic. Such spatial distribution of dust was observed by space probes Helios 1 and 2 at distances from 0.3 to 1 Astronomical Units (AU) from the Sun \cite{PS}. It was found that the dust concentration decreases in ecliptic as a function of $r^{-1/3}$, and at the distance $r\geq3$ AU, dust is almost absent \cite{LRP}. So for the further estimates, we can take $R_*=3$ AU as a critical radius. The total mass of dust in the Solar system estimates as $M_d=10^{19}-10^{20}$ g, and the major part of the mass (about 2/3 of the total mass), is concentrated in the dust particles of mass $m_d$ in the mass range ${10}^{-3}<m_d<{10}^{-5}$ grams \cite{PS}. Using these data, we can obtain an estimate of the average density for the dust component in terms of the hydrogen atoms in a sphere of a radius $R_*$: $n_d=3M_d/(4\pi R_*^3\mu_H)={10}^3$ cm$^{-3}$, here $\mu_H=1.66\cdot 10^{-24}$ grams is the mass of the hydrogen atom.

As is known, neutral gas in the Solar system was discovered by observing
resonantly scattered solar radiation. At distances of $r=5$ AU from the Sun, the neutral gas is distributed practically uniformly or homogeneous at a concentration of hydrogen atoms on the average being 0.06 cm$^{-3}$, helium 0.012 cm$^{-3}$, and the gas temperature is close to 9000 K \cite{PS,Bei}. Near the Sun, the distribution of gas becomes highly heterogeneous due to the influence of solar attractions such as: Solar flares - a large explosion in the sun's atmosphere, Coronal mass ejection - a massive burst of solar wind sometimes associated with solar flares, Geomagnetic storms - the interaction of the Sun's outburst with Earth's magnetic field, Solar proton events or proton storms, ultraviolet radiation and finally the solar wind. In cosmic rays, a component containing partially ionized atoms of helium, oxygen, nitrogen, and neon is also observed. In this component, particles with a specific energy of 1-100 MeV/nucleon are also noticed; their flux increases with distance from the Sun and the plane of the solar equator \cite{PS,Bei}. The particles of an anomalous component are assumed to acquire high energy at the heliosphere limit, in the area of interaction of the solar wind with interplanetary plasma. As is seen, the density of this component of the interplanetary medium is much smaller than the contribution of the solar wind and dust and it can be neglected.

Thus, proceeding from these data, we come to the conclusion that only the matter of solar wind and interplanetary dust are really suitable as a fuel for plasma thrusters. For further estimates, we shall set that the density of the interplanetary medium $n$ varies in the range 
$n_w \leq n \leq n_d$, i.e. from 10 to $10^3$ cm$^{-3}$.

In order to understand the possibility of using solar radiation for ionizing neutral atoms and their subsequent acceleration, we estimate the radiation power distribution as a function of the distance from the Sun. It should be noted that here we limited our consideration to the distances on the boundary of the asteroid belt. Taking into account that the surface temperature of the Sun at a distance of $R_S={10}^8$ m, is $T_S$=6000 K \cite{PS,Bei}, from the Stephan-Boltzmann's law we get
\begin{equation} 
W_s=\frac{\sigma T_S^4}{\left(r/R_s\right)^2}\/,
\label{fl_2}
\end{equation}
where $\sigma$ is the Stefan-Boltzmann constant. This dependence is sketched in Fig. \ref{fig1}, where for the convenience of perception, the distances from the Sun to the planets falling into a sphere of radius $R_*$ are labeled.
\begin{center}
\includegraphics[width=9cm, height=6cm]{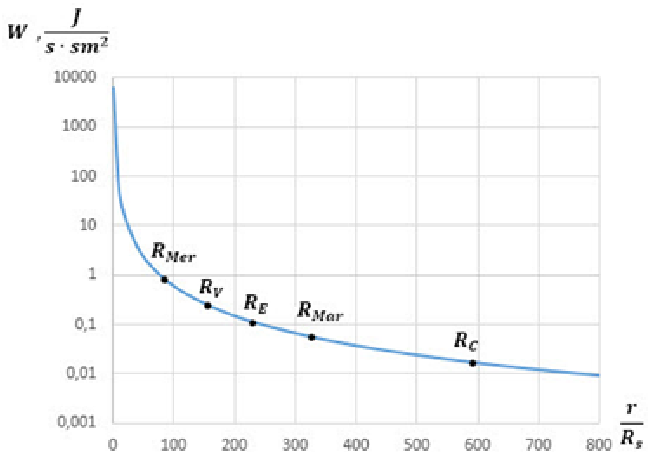}
\end{center}
\refstepcounter{graf}\label{fig1}
\hspace*{25mm}\parbox[b]{9cm}{\small {\bf Fig. \thegraf.} \hspace*{1mm} The radiation power with respect to the distance from the Sun. Here $R_{Mer}$ corresponds to the Mercury orbit, $R_V$ - the Venus orbit, $R_E$ - the Earth orbit, $R_{Mar}$ - the Mars orbit and $R_C$ - the Ceres orbit.}\\
\section{The plasma production from interplanetary medium}

Proceeding from information about the distribution of interplanetary medium and the Solar radiation, we can estimate the parameters of plasma where we can produce an interplanetary medium in a plasma thruster as a fuel. Also, we have discussed the possible ways to increase the plasma density up to generate thrust at minimally required values.

As a plasma source, we shall consider the cylindrical charged chamber of volume $V=10^3$ cm$^{3}$ where a low-pressure discharge is induced. As a rule, such types of discharges can be performed with the help of an HF microwave field at a pressure of $p \sim 10^{-3}$ Torr (see, for example, \cite{GK,Rz}) that corresponds to the particle density $n_0 \sim {10}^{10}-{10}^{11}$ cm$^{-3}$ under normal conditions. This indicates that it is impossible directly to use the interplanetary medium to organize a low-pressure discharge. We need to apply an additional technical parameter to obtain the required number of particles $N=n_0 V=10^{14}-10^{15}$ into the chamber under consideration.
\begin{center}
\includegraphics[width=9cm, height=5cm]{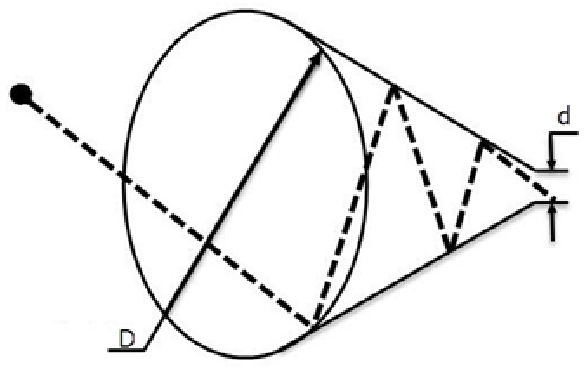}
\end{center}
\refstepcounter{graf}\label{fig2}
\hspace*{25mm}\parbox[b]{9cm}{\small {\bf Fig. \thegraf.} \hspace*{1mm} The scheme of trap for interplanetary substance (here, we set D=200 cm and d=20 cm).}\\
We presume to do this with the help to trap interplanetary substances by using a special geometry (see Fig. \ref{fig2}) and the accumulation of particles with increasing the spacecraft's flow velocity with respect toward incoming particles. Such a conical structure should be placed on a moving spacecraft in the direction of its travel to minimize pseudo-drag from the momentum of the particles acting against the inlet scoop. We will consider this effect as inconsequential. Furthermore, the surface of such conical construction is suitable for the solar panel location.

Let us explain the simple mechanism of particle capture by assuming that the particle medium is stationary. In this case, the flux of particles that reach the surface, due to the relative velocity of the particles relative to the ship, is determined only by the ship velocity. The falling particles, repeatedly are reflected from the conical surface and fall into the region of the discharge chamber. Proceeding from the continuity of the particle flux: $nD^2=n_cd^2$, where $D$ is the diameter of the inlet, $d$ is the input diameter of the discharge chamber. We see that the density of particles in the narrow part of the conical funnel $n_c$ can be increased with respect to the input density $n$ in $k =D^2/d^2$ times. For example, for the sizes $D$ and $d$ of the trap shown in Fig.\ref{fig2}, we obtain $k=10^4$. It should be borne in mind that this variant of density enhancement is workable as long as the medium
remains collisionless, i.e. pressure effect does not work as previously assumed. In our case, the collisionless regime may be violated in the narrowest part of the trap. So in estimations of the mean free path $\lambda$ we put the density in the discharge chamber being equal to the desired value $n_0=10^{11}$ cm$^{-3}$. Then for the characteristic gas-kinetic cross-section $\sigma_g=10^{-16}$ cm$^2$, we get $\lambda=1/(\sigma_g n_0)=10^5$ cm that is much larger than the characteristic size of the discharge chamber, so there is no need to take into account the initial pressure of the medium.

\begin{center}
\includegraphics[width=9cm, height=6cm]{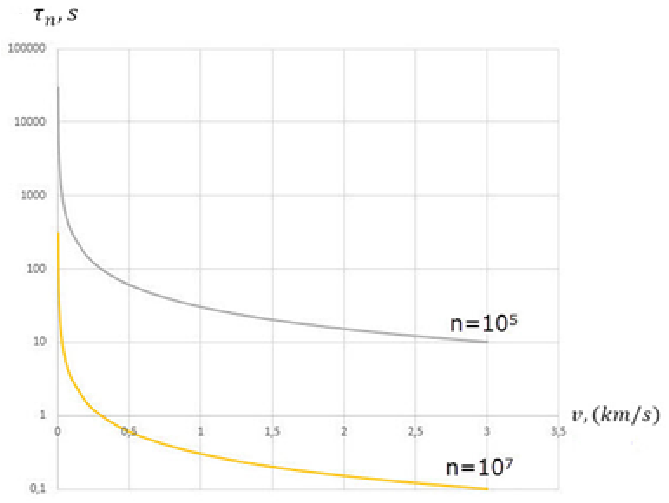}
\end{center}
\refstepcounter{graf}\label{fig3}
\hspace*{25mm}\parbox[b]{10cm}{\small {\bf Fig. \thegraf.} \hspace*{1mm} Accumulation time as a function of the relative velocity for the incoming flow of the solar wind particles ($n_c=10^5$  cm$^{-3}$) and the dust particles ($n_c=10^7$  cm$^{-3}$).}\\

Now we shall pass over to estimate the accumulation of particles due to a change in the relative velocity of the interplanetary medium. It is easy to see that the flux of the interplanetary particle incident on the trap has filled the selected volume to a given number of particles $N$ for the time $\tau_n$:
\begin{equation} 
\tau_n=\frac{4N}{\pi d^2v n_c},
\label{fl_3}
\end{equation}
where $v$ is the relative velocity of incoming flow, $n_c=k\cdot n$ is incoming particles after passing the trap with $k=10^4$ the input radius of the discharge chamber. Here we restrict our consideration by two characteristic cases: the flux is created by the solar wind with density $n=10$\ cm$^{-3}$ and the flux consists from dust particles with a density of $n=10^3$ cm$^{-3}$. In Fig. \ref{fig3} the characteristic times of particles accumulation for these both cases are plotted as a function of the relative velocity, for which we take the ship
velocity.

To assess the necessary energy expended on the ionization of neutral atoms and their acceleration, we shall use the data on the radiation power for the Sun presented in Fig. \ref{fig1}. In these estimates, we assume that the ionized medium consists only of hydrogen atoms having a minimum ionization potential of $\epsilon_i=13.6$ eV. For simplicity, we shall neglect to brake the ship by the oncoming flux of interplanetary medium and restrict our consideration by only elastic head-on collisions of the oncoming particles with the spacecraft surface. Despite the obvious physical fallacy of such an assumption, the result implies this is the simplest possible way to account for the contribution of particle
deceleration in the energy balance:
\begin{equation} 
W S_* \tau_{\epsilon} = \epsilon_i N+\mu_H N \left(\frac{v^2}{2}+\frac{v_{ex}^2}{2}\right),
\label{fl_4}
\end{equation}
where $\tau_{\epsilon}$ is the accumulating time for ionization and accelerating particles by the sun radiation, $S_*$ is the square of a solar panel. As previously mentioned, the elements for a solar battery may be located on the surface of the trap so in the present estimates one can use $S_*=\pi D^2/4$.

The value of velocity $v_{ex}\geq 0$ is determined by the acceleration method and an additional discussion of this moment is required, but in order to obtain a minimum energy estimation in order of magnitude one can assume $v_{ex}=0$. Then from (\ref{fl_4}), we get 
\begin{equation} 
\tau_{\epsilon}=\frac{\epsilon_iN + \mu_H N v^2/2}{W(r)\pi D^2/4}\/.
\label{fl_5}
\end{equation}
Depending on the velocity of the device, the power of solar radiation and the geometric sizes of a solar panel, this relation determines the characteristic time for the required energy accumulation. However, it should be borne in mind that these estimates of the times $\tau_{\epsilon}$ should be considered as upper-bound marginal valuations.

Fig. \ref{fig4} presents the dependence $\tau_{\epsilon}$ for the space region between Venus and the asteroid belt between Mars and Jupiter. In these estimations, we take $S_*=80$ m$^2$. As is shown from this curve, the value $\tau_{\epsilon}$ changes from 0.1 s at a distance $\sim 1$ AU
to several seconds for distances $r \geq 6$ AU. Under such conditions, it is impossible to maintain a continuous operation mode of the plasma thruster but a pulsed engine might be workable. Comparison of the dependencies presented in Figs. \ref{fig3} and \ref{fig4} show that the magnitude of pulse primarily depends on the capturing interplanetary matter that is brought about by the ship's velocity with respect to the interplanetary substance. One can try to reduce the pulse time by increasing the area of the trap, however, this will affect the dimensions of the spacecraft. Increasing the craft's velocity makes it possible to decrease the accumulation of time. In this case, a continuous mode of operation is not ruled out for high velocities.
\begin{center}
\includegraphics[width=9cm, height=6cm]{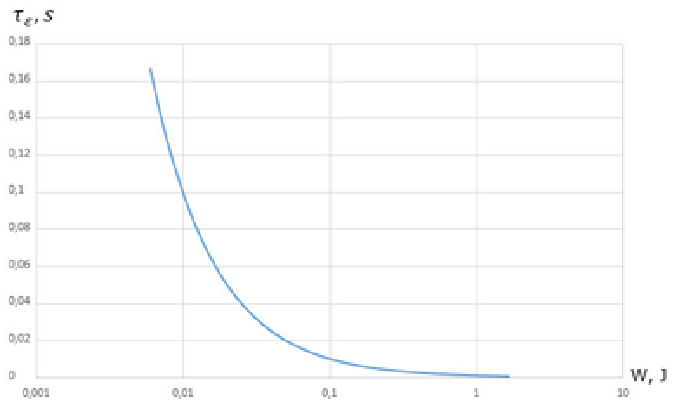}
\end{center}
\refstepcounter{graf}\label{fig4}
\hspace*{25mm}\parbox[b]{9cm}{\small {\bf Fig. \thegraf.} \hspace*{1mm} The value of the energy accumulation time as a function of $W(r/R_S)$ under $S_*=80$ m$^2$.}

In the present model, we have studied the use of the interplanetary substance as a fuel for plasma engines. For this, we used astrophysical data on the density of the cosmic medium, which we recounted relative to hydrogen atoms. It appears that the minimum and maximum density in terms of hydrogen atoms varies in the range from $n_{min}=10$ to $n_{max}={10}^3$ ${cm}^{-3}$. The minimum estimation corresponds to taking into account only the solar wind and the maximum value that is obtained from interplanetary dust concentrated within a sphere of radius 3 AU from the Sun. Moreover, a low-pressure plasma source requires a density of the order $n_c\geq 10^7$ cm$^{-3}$. This means that for the low velocities of spacecraft, it is impossible to use the interplanetary medium as a fuel for plasma thrusters. However, with increasing rocket's velocity, as expected, the use of additional capturing space particles makes it possible to increase the density of the discharge volume. So in the example considered, the trap increases the particle density in $k = 10^4$ times irrespective of the rocket velocity only due to the ratio of geometric sizes. The dependencies presented in Fig. \ref{fig3} indicates that the accumulation time of particles for the accepted value $n_c$ depends on the device velocity. It should be noted that for simplicity, in the present estimations we have neglected by the relative velocity of particles with respect to the rocket. In the general case, this is not justified. As seen from the above dependencies, the incoming particle flux can only provide a pulsed mode of operation for the plasma source. In this case, the pulse frequency of such a device must vary with the rocket velocity and it depends on the density of the external medium.

\section{Some speculative considerations}

From these results, it follows that the basic issue for plasma thrusters in space where $r\leq5-6$ AU, there is the problem of getting enough substance which may be used as propellant for plasma production. Taking into account the nonzero strength at every point of space for the interstellar magnetic and the gravitational fields, one can suggest a way of additional extracting the interplanetary plasma medium with the help of drift plasma in crossed magnetic and gravitational fields. Indeed, if these fields are not collinear they bring about the drift of
electrons and ions with velocities:
\begin{equation} 
{\bf V}_s=\frac{{\bf F}_s \times {\bf B}_{ism}}{q_s B_{ism}^2} c,
\label{fl_6}
\end{equation}
here the label $s=e$ relates to the electrons and the subscript label
$s=i$ relates to the ions, $q_s$ is the charge of s-component, ${\bf F}_s$ is the gravitational force acting on a particle of mass $m_s:$
$$F_s=-Gm_s\int \frac{\rho(t,{\bf r}^{\prime})({\bf r}-{\bf r}^{\prime})}
{\vert {\bf r}-{\bf r}^{\prime}\vert^3}dV^{\prime}\/,$$
where $G$ is the gravitational constant, and $\rho(t,{\bf r})$ is the distribution of matter density in the solar system. Since $m_e
\ll m_i$ then the drift mass flux is determined by the ion component

\begin{equation} 
{\bf j} = \frac{G c q_i m_i n}{B_{ism}^2}{\bf B}_{ism}\times\int\frac{\rho(t,{\bf r}^{\prime})({\bf r}-{\bf r}^{\prime})}{\vert {\bf r}-{\bf r}^{\prime} \vert^3}dV^{\prime}.
\label{fl_7}
\end{equation}
This relation shows that in the solar system there exist some points where the value of ${\bf j}$ may be very large and one can try to choose the spacecraft trajectory to extract this additional substance.

Note that this does not find or include any 'Dark' matter which, if it existed, might be suitable as a propellant.  However, the current wisdom claims that dark matter has neither electrical nor magnetic properties, thus this would not be suitable as a propellant in a plasma thruster unless some yet to be determined process can be obtained for ionization.

The viewpoint is that with the mass flow rate ingested from the inlet, either magnetic/electrical or in a benign fashion would define the capture cross-section of these particles. As seen, the density diminishes with distance from the Sun. The only other variable is the velocity. Thus to increase velocity, the flow rate can be increased to operate. This can become prohibitive and may imply that the system can only initially operate within the near-Earth environment before it requires reaching some acceptable velocity, say .2 c or higher before these conditions could operate. This would definitely push contemporary technology.

As suggested, a pulsed configuration may be suitable simply because of the low pulse frequency due to the density distribution at long ranges. Another issue is to reclaim the propellant and reuse it to increase the flow rate distribution. If a magnetic field is employed, the thrust occurs as the particles in the plasma thruster leave the spacecraft in a selected direction but as these particles follow the lines of force as a loop, they reach the nose of the spacecraft and the net thrust obtained then goes to zero. One alternative is to use an intermittent magnetic field so that the plasma may generate thrust when the field ceases and reclaims the mass when the field restarts in a continuous cyclic high-frequency fashion. 

An alternative with limitations to the amount of propellant available would be a propellant concept similar to the Woodward inertia drive. Here, mass is placed in an electric/magnetic field that uses a current to generate inertial motion. To restart the process, the fields are used to increase the time period for the movement of inertia in the original position and this gains a net thrust albeit at very low levels with current technology. If we look at the closed magnetic lines of force methodology previously mentioned, the approach may be to increase particle velocity axially from the thruster and as the field motion is radial, to decelerate it so that the high velocity diminishes and the particles move in the outer loop at a lower speed in the lines of forces moving up toward the nose of the spacecraft and then radially decelerated more before being reinjected into the plasma thruster. Such a magnetic field, for example, may be representative to a cone. Of course, this is speculative but may offer an interesting inertial propulsion concept possibility.

Another approach may be available by using a Poynting drive with electric and magnetic fields that would be independent of exterior propellant needs. Here, thrust is obtained directly most likely in a closed system. Moreover, a propulsion concept can include an electrodynamic tether in which electrons are collected from the ambient plasma at one end of conducting strand and then emitted at the other end of the tether to produce thrust (see, for example \cite{Joh,FWJ}). These types of concepts may be the only reasonable alternatives to reach interstellar travel because of limitations due to available propellant either onboard or by drawing ions from the exterior environment.

\section{Conclusion}

There are other considerations worthy of investigating this discipline. Interestingly, the solar radiation is considered as an energy source for the ionization of accumulated particles and their further acceleration. Using the power distribution of the solar radiation, we have estimated the time needed to accumulate energy for ionization and acceleration to the required rocket velocity (see Fig. \ref{fig4}). As seen from the comparison of these dependencies with the accumulation time of particles, the main time will be spent on the particles accumulation regardless of the rocket velocity. In this case, the values for the accumulation time of particles and energy show that the space medium can be used as a fuel for a plasma source operating in the pulsed mode for flights in a radius of 3 AU from the Sun, even with existing technical capabilities. This may also occur for low particle values at the Ceres, where the time particles are hit for collection, the system may automatically operate like a pulsed system. The implication that a pulsed system with this requirement may be more profitable than a continuously operating system which would end up with propellant starvation.

In this regard, we dwell in brief on the features of propelling a plasma discharge. It is clear that at low pressures, it is difficult to use a high-frequency discharge since it is caused by the particle collisions which occur rarely. In this case, there appears to be a problem how one should handle a required pulsed system of the discharge system. Microwave discharges in a waveguide may be used to enhance performance for the plasma production or the initiation of a discharge in a low-pressure gas where it may be propelled by microwave radiation with a stochastically jumping phase (see, for example, \cite{Rz,Chen,LL}). Although the first option is more technically simple as using microwaves as a breakdown power minimum value for creating a stochastically jumping phase that depends weakly on a working gas pressure caused by the anomalous nature of collisionless electron heating \cite{Chen}. It should be borne in mind that such collisionless heating may come about in some non-local regime when there are appreciable spatial gradients in the fields, for example, the electric field can be increased near the boundaries. In fact, non-Ohmic electron heating can occur in the presence of an oscillatory current crossing a plasma boundary. Such a feature allows extending the discharge existence region in a direction of low pressures by using some special geometry. Therefore, it would be useful to combine both methods in examining this effect on the source.

For this model, there is a large influence of astrophysical data that implies at large distances, this concept may be questionable in a continuous fashion.  We have addressed the problem by investigating a pulsed system and here, limitations can be factored to succeed by increasing the spacecraft's velocity. However, this may be pressing considering contemporary technology. Further, the issue of reusing the plasma by incorporating a magnetic field loop, although speculative, may also be promising as well and that this concept of extracting ions from the space environment may not be promising at long range missions. Concluding comments imply that a pulsed system may operate or at interstellar distances, extraction of external propellant may demand other means for generating propulsion. These other concepts would push the current state-of-the-art.

\end{document}